\documentclass[aps,prl,twocolumn,groupedaddress,showpacs,
preprintnumbers]{revtex4}

\usepackage{graphicx}
\usepackage{dcolumn}
\usepackage{bm}

\begin{document}
\preprint{IRB-TH-2/06}

\title{NLO perturbative QCD predictions for 
$\gamma \gamma \to M^+ M^-$ ($M= \pi ,\,K$)}



\author{Goran Duplan\v ci\' c}
\email{gorand@thphys.irb.hr}
\author{Bene Ni\v zi\' c}
\email{nizic@thphys.irb.hr}
\affiliation{Theoretical Physics Division,
Rudjer Boskovic Institute,
P.O.Box 180,
HR-10002 Zagreb,
Croatia
}


\date{\today}

\begin{abstract}
We report the first complete leading-twist 
next-to-leading-order perturbative QCD
predictions for the two-photon exclusive channels
 $\gamma \gamma \to M^+ M^-$ ($M= \pi ,\,K$) 
at large momentum transfer.
The asymptotic distribution amplitude is utilized as a candidate
form for the nonperturbative dynamical input.
Comparison of the obtained results with the existing experimental data 
does not provide sufficiently clear evidence to support the applicability 
of the hard-scattering approach at currently accessible energies.
\end{abstract}


\maketitle

Exclusive processes at large momentum transfer
provide a particularly suitable testing ground
for perturbative QCD (PQCD).
In the hard-scattering approach (HSA) \cite{formalizam}
the amplitudes of these processes are given in terms of a convolution
of a hard-scattering amplitude of collinear hadron constituents, and 
distribution amplitudes summarizing soft physics.
Although the HSA is widely used, its applicability
at currently accessible energies is still being questioned.

The leading-order (LO) PQCD predictions have been
obtained for many exclusive processes, but owing to their 
sensitivity to the choice of the renormalization scale
and scheme, they do not have much predictive power.
To stabilize the LO results, and to achieve a
complete confrontation between theoretical
predictions and experimental data, it is crucial
to obtain higher-order contributions.
However, owing to great technical difficulties involved,
the complete next-to-leading-order (NLO) predictions were 
obtained only for
the pion electromagnetic form factor \cite{elff,ff} and the pion
transition form factor \cite{ff,tff}.

The two-photon annihilation into flavor-nonsinglet
helicity-zero mesons,
 $\gamma \gamma \to M^+ M^-$
 ($M= \pi ,\,K$),
is one of the simplest hadronic processes.
Owing to the pointlike structure of the photon,
the initial state is simple, with
strong interactions present only in the final state.
The energy behavior, angular behavior, and normalization of
the cross section of this process are nontrivial predictions
of PQCD.
This process has been the subject of a number of 
experimental investigations. The most recent data
(DELPHI \cite{DELPHI},
ALEPH \cite{ALEPH}, and Belle \cite{Belle})
are obtained
for the center-of-mass energies $W <6 \,{\rm GeV}$
and the scattering angles in the range $|\cos \theta |<0.6$.

In the HSA, at large momentum transfer
(high $W$ and large $\theta$), the process is 
described by the helicitiy amplitude
\begin{eqnarray}
{\cal M}(\lambda \lambda';W,\theta )&=& 
\int_0^1 \!\!\! dx \int_0^1 \!\!\! dy 
\,\, \Phi_M^*(x,\mu_F^2)\, \Phi_M^*(y,\mu_F^2) \nonumber
\\
& &{}\times T_H(\lambda \lambda';x,y;W ,\theta;\mu_F^2,\mu_R^2),
\label{eq:1}
\end{eqnarray}
where $\lambda$ and $\lambda'$ are photon helicities,
$\mu_F^2$ is the factorization
and $\mu_R^2$ the renormalization scale.

The function $\Phi_M(x,\mu_F^2)$ is the meson distribution amplitude
(DA), representing the amplitude for the meson consisting
of a $q\overline{q}$ pair, with the quark and antiquark
collinear and on shell, and sharing fractions $x$ and $1-x$
of the meson's total momentum.
There is a long ongoing debate in the literature \cite{asDA} regarding 
the 
correct form of the DA at currently accessible energies. 
For the purpose of this work we approximate the DA by its asymptotic 
form and choose 
$\Phi_M(x,\mu_F^2)=\Phi_M^{as}(x) \equiv
f_M \sqrt{3/2}\, x(1-x)$,
where $f_M$ is the meson decay constant
($f_{\pi\,(K)}=0.131\,(0.160)\,{\rm GeV}$).

The function $T_H(\lambda \lambda';x,y;W,\theta;\mu_F^2,\mu_R^2)$
in Eq. (\ref{eq:1}) is the hard-scattering
amplitude for producing collinear meson
constituents from the initial photon
pair, is free of collinear singularities, and has 
a well-defined perturbative expansion.

\begin{figure}
\includegraphics{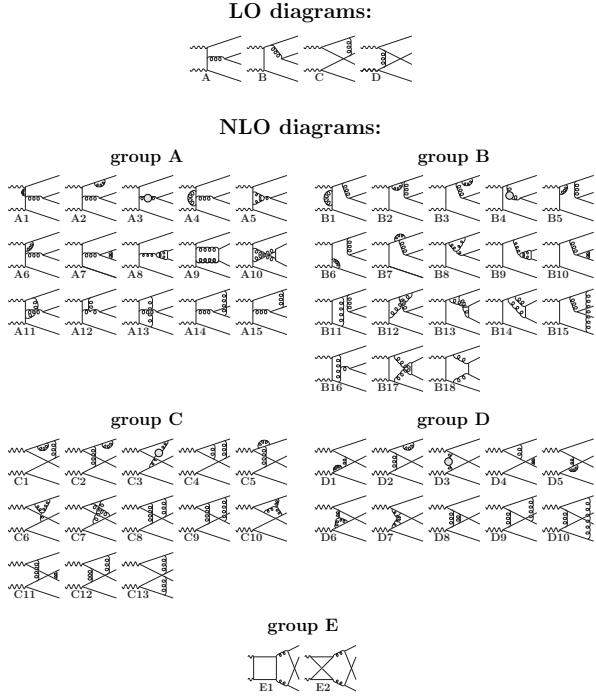}
\caption{\label{figa}
Distinct diagrams 
contributing to the
$\gamma \gamma
\to (q \overline{q})+(q \overline{q})$
amplitude at the NLO. The total number of LO diagrams is 20, and
the total number of NLO diagrams in groups A, B, C, D, and E
are 96, 156, 94, 70, and 6, respectively.}
\end{figure}

At the NLO the helicity amplitude in Eq. (\ref{eq:1})
is given by the NLO result for $T_H$,
obtainable from the LO and the NLO contributions to the
$\gamma \gamma
\to (q \overline{q})+(q \overline{q})$
amplitude,
with massless quarks collinear with outgoing mesons.
These contributions arise from 20 tree and 422 one-loop 
Feynman diagrams, respectively.
Distinct diagrams are shown in Fig. \ref{figa}.
The other diagrams are obtained from these
using various symmetries.

The LO prediction for the process
$\gamma \gamma \to M^+ M^-$
 ($M= \pi ,\,K$)
is due to Brodsky and Lepage (BL) \cite{LOffMM}.
Making use of the similarity of the
LO expressions of the helicity amplitudes for the
process in question and
the timelike meson electromagnetic form factor $F_M$,
they obtained the approximate relation
\begin{equation}
\frac{{\rm d}\,\sigma}{{\rm d}\,\cos \theta} (\gamma\gamma \to
M^+ M^-)
\approx \frac{8 \pi \alpha^2}{W^2}\frac{|F_M(W^2)|^2}{\sin^4
\theta}. \label{ffBL}
\end{equation}
Despite the fact that the complete LO prediction is given in Ref. 
\cite{LOffMM}, the approximate relation
of Eq. (\ref{ffBL}) is what is usually referred as the Brodsky-Lepage
prediction. 

The first attempt to perform the
NLO analysis was done by Nizic \cite{bene}.
Lacking powerful analytical methods
for calculating one-loop Feynman diagrams
with five and six internal lines, 
the NLO prediction in Ref. \cite{bene} was obtained
with the simplest equipartition meson DA
$\Phi_M \propto \delta (x-1/2)$.
Also, the contribution of the diagrams of
group E was omitted. Therefore, the prediction
of Ref. \cite{bene} cannot be considered complete.

Here we present the results of a complete
NLO analysis of the process
$\gamma \gamma \to M^+ M^-$ ($M= \pi ,\,K$).
Using the method of Refs. \cite{algoritam}, we
evaluated all the one-loop diagrams analytically.
It should be pointed out that the computation
of one-loop diagrams with five and six
internal lines represents a highly demanding task.
The details will be given elsewhere.

Our complete NLO PQCD result for the cross section,
in the angular range 
$|\cos \theta |<0.6$,
assuming the asymptotic DA and using the 
$\overline{MS}$ renormalization scheme, is
\begin{eqnarray}
\sigma (\gamma\gamma \to \!M^+ M^- ) &=&
f_M^4 \, \frac{1.035}{W^6}\,
\alpha_{\mbox{\tiny S}}^2(\mu^2_R) \nonumber \\
& &{}\hspace{-3.5cm} \times \left\{ 1+
\frac{\alpha_{\mbox{\tiny S}}(\mu^2_R)}{\pi}
\left[ -3.828+
\frac{\beta_0}{2}\left(3.563+ \ln \frac{\mu^2_R}{W^2}\right)
\right]
\right\}\!\!,\label{final}
\end{eqnarray}
where
$\alpha_{\mbox{\tiny S}}(\mu_R^2)= 
4 \pi/ (\,\beta_0 \ln(\mu_R^2/\Lambda^2)\,)$
is the coupling constant
renormalized at $\mu_R^2$
(function of $W^2$ only), with
$\beta_0=11-2n_f/3$ ($n_f=3$)
and $\Lambda=0.2\,{\rm GeV}$.
As a consequence of using the asymptotic DA, 
the result (\ref{final}) does not depend on $\mu_F^2$.
The corresponding result for the differential
cross section is too lengthy to be presented here.

As it is seen from Eq. (\ref{final}), the NLO corrections,
at accessible energies ($W<6\,{\rm GeV}$), are substantial.
Thus, for $\mu_R^2=W^2$ (pragmatical choice),
the ratio of the NLO to the LO contributions is 0.9
for $W=4\,{\rm GeV}$, where $\alpha_{ \mbox{\tiny
S}}(\mu_R^2)=0.23$, and becomes
less than 0.5 only for $W>45\,{\rm
GeV}$, 
where $\alpha_{ \mbox{\tiny S}}(\mu_R^2)=0.13$.
Obviously, the problem of convergence relates to
the size of the NLO coefficient,
not to the coupling constant.

It is important to observe that the large contribution to the
NLO coefficient in Eq. (\ref{final}) comes from the $\beta_0$ 
term, arising
from the vacuum-polarization diagrams, and it is desirable
to resum them into the running coupling constant.
This can be achieved using the Brodsky-Lepage-Mackenzie
(BLM) scale setting procedure \cite{BLM}.

However, the fact that the form of the effective coupling 
at low energies, as well as its behavior at timelike
scales are not determined, together with the fact that
the process at hand contains both spacelike and timelike
BLM scales, does not allow one to perform the BLM
procedure in a completely satisfactory way.
For the sake of clarity, 
we discuss only the BLM improved LO prediction
LO(BLM), which is of the broadest interest.
The BLM scales were determined for each
LO diagram separately, for different helicities, angles,
and quark momenta. The mean-value theorem for integration was used to
avoid too low scales in the integration indicated in 
Eq. (\ref{eq:1}).
To deal with the timelike scales, the continuation
$\ln(\mu^2/\Lambda^2)\to \ln(\mu^2/
\Lambda^2)-{\rm i} \pi$
was used in the
coupling constant.

\begin{figure}
\includegraphics{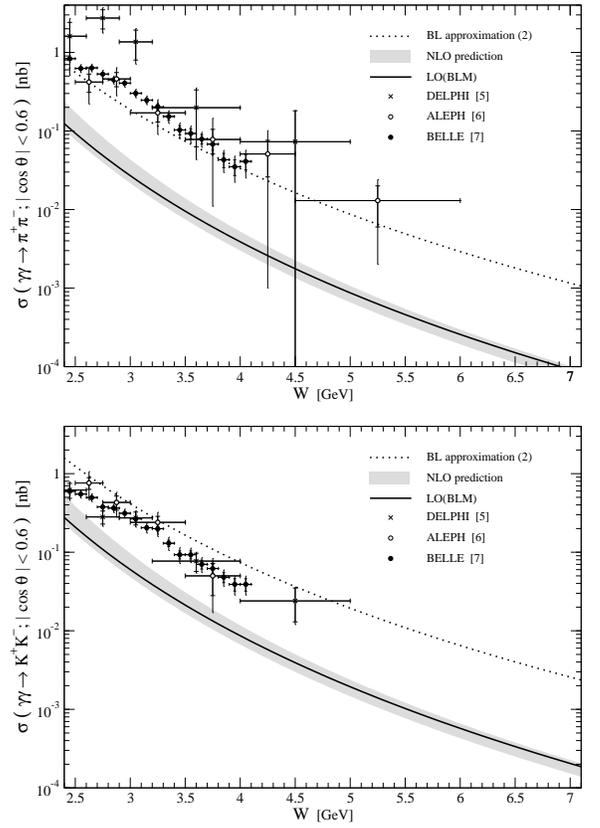}%
\caption{\label{figpion} Cross sections for the processes 
$\gamma \gamma \to M^+ M^-$.}
\end{figure}

\begin{figure}
\includegraphics{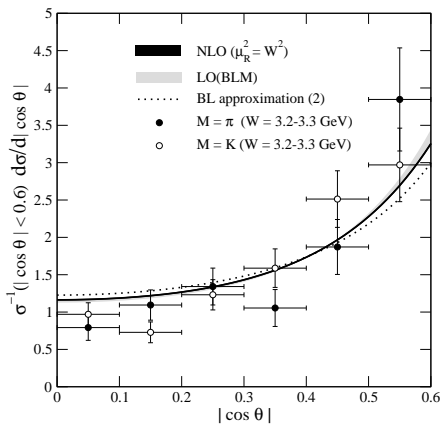}%
\caption{\label{figkut}  Angular dependence
$\sigma^{-1}\, \,{\rm d}\sigma/{\rm d}|\cos \theta
|$ for the $\gamma\gamma \to M^+ M^-$ processes.
The experimental points are from Ref. \cite{Belle}.}
\end{figure}

In Figs. \ref{figpion} and \ref{figkut} 
we compare our predictions with 
the recent experimental data and the BL estimate in Eq. (\ref{ffBL}).
To assess the theoretical uncertainty related to the
choice of the scale $\mu_R^2$ in Eq. (\ref{final}), in Fig. 
\ref{figpion} we show
the range of the NLO predictions (shaded region)
obtained by varying  $\mu_R^2$ from
$\mu_R^2=W^2$ down to the scale where the NLO prediction
reaches maximum, $\mu_R^2\approx W^2/15$  (the
scale determined by the fastest-apparent-convergence
principle \cite{FAC}).
The NLO prediction (not shown in Fig. \ref{figpion}) for
$\mu_R^2\approx W^2/20$, the scale set by
the principle of minimal sensitivity \cite{PMS}, is located 
slightly below the upper limit.
The solid line in Fig. \ref{figpion} is our LO(BLM) prediction,
while the dotted line is the BL estimate based on Eq. (\ref{ffBL})
with the form factors approximated by
$F_{\pi}(W^2)\approx 0.4\,{\rm GeV}^2/W^2$
and $F_{K}(W^2)\approx f^2_K/f^2_{\pi}\,
F_{\pi}(W^2)= 0.6\,{\rm GeV}^2/W^2$,
as in the original paper \cite{LOffMM}. 
However, the recent
measurements \cite{tlff}, suggesting that
$F_{\pi\,(K)}(W^2)\approx 1.01\,(0.85)\,{\rm
GeV}^2/W^2$ (around $W^2= 13.48\,{\rm GeV}^2$)
put the success
of the widely cited BL prediction (dotted line)
to question.
Figure \ref{figpion} shows that our results for the cross section
are still significantly, but less than an order of
magnitude, below the data.

Next consider the $W$ dependence of the cross section.
Using the parametrization
$\sigma (\gamma\gamma \!\!\to \!M^+ M^- )\propto W^{-n_M}$
the results obtained by Belle \cite{Belle} give
$n_{\pi}=7.9\pm 0.4\pm 1.5$ and 
$n_{K}=7.3\pm 0.3\pm 1.5$, for
$3.0\,{\rm GeV} < W < 4.1\,{\rm GeV}$.
On the other hand,
we find from Eq. (\ref{final}) that the power $n_{\pi}(=n_K)$ 
takes the values
6.9(7.4) for $\mu_R^2=W^2 (W^2/15)$.
The LO(BLM) and BL predictions are 6.7 and 6, 
respectively.
As for the ratio of kaon to pion cross sections,
our prediction equals $f^4_K/f^4_{\pi}=2.23$, coinciding
with the BL prediction, while the Belle results \cite{Belle}
for $3.0\,{\rm GeV} < W < 4.1\,{\rm GeV}$
give $0.89\pm 0.04\pm 0.15$.

In Fig. \ref{figkut} we show the angular dependence of the
ratio $\sigma^{-1}\,\,{\rm d}\sigma/{\rm d}|\cos \theta|$.
The range of the NLO and LO(BLM) predictions
is obtained by varying 
$W$ from 2.4 GeV to 4.1 GeV (range
covered by the Belle experiment \cite{Belle}).
The BL curve, based on Eq. (\ref{ffBL}), corresponds to
$1.227/\sin^4\theta$.
To keep figure synoptic,
experimental points are given only for 
$W=3.2-3.3\,{\rm GeV}$ (middle of the Belle range \cite{Belle}).

As it is seen from Figs. \ref{figpion} and \ref{figkut}, the 
energy and the angular dependence of our NLO and LO(BLM) predictions
for the cross section are in very good agreement with the 
data. As for the absolute normalization,
our results are still significantly below the data.
Obviously, the normalization represents a problem.

This problem has been the subject of a long debate,
and critics use it as an argument
to discard, at experimentally available energies,
either the HSA \cite{kroll} 
or the asymptotic DA as an appropriate distribution amplitude
\cite{Cher}.
The strength of criticism depends on the size
of higher-order corrections.
Here we have shown that the NLO corrections can be substantial,
so a part of the normalization problem can be attributed to the
size of uncalculated higher-order corrections.
However, assuming 
reasonable scale variation of the
NLO prediction, or controlled perturbative series (LO $>$ NLO term $>$ NNLO
term)
it is not to be expected that inclusion of higher-order corrections
would lead to a solution of the normalization problem.
On the other hand, 
good agreement between theory
and experiment for angular dependence (Fig. \ref{figkut}), and the
disagreement for the ratio of the kaon to pion cross sections
suggest that the specific form of the DA certainly plays a relevant role.
An analysis of various choices of the DA, however,
is outside the scope of this letter.

As a quantity which is expected to be largely
insensitive to both the form of the
effective coupling as well as to the choice of the DA,
and as such representing a nontrivial PQCD prediction,
we next consider the ratio of 
$\sigma (\gamma\gamma \to M^+ M^-)$ and
$|F_M(W^2)|^2$ at the NLO.

The result of our NLO calculation of the timelike
meson form factor is of the form
\begin{eqnarray}
|F_M (W^2)|^2 &=&
f_M^4\, \frac{64 \pi^2}{W^4}\,
\alpha_{\mbox{\tiny S}}^2(\mu^2_R)\nonumber \\
& &{}\hspace{-2.5cm} \times \left\{ 1+\frac{
\alpha_{\mbox{\tiny S}}(\mu^2_R)}{\pi} \left[
-7.833+\frac{\beta_0}{2}\left(\frac{14}{3}+ \ln
\frac{\mu^2_R}{W^2}\right)\right]\right\}\!,\label{finalff}
\end{eqnarray}
and is consistent with the NLO result for
the spacelike form factor \cite{elff}.
On the basis of Eqs. (\ref{final}) and (\ref{finalff}),
and using the same $\mu_R^2$, we find
\begin{equation}
W^2\frac{\sigma (\gamma\gamma \!\!\to \!M^+\! M^-
)}{|F_M(W^2)|^2}=638 \left[1-\alpha_{\mbox{\tiny S}}(\mu^2_R)\,0.306
\right]{\rm nb}\,{\rm GeV}^2. \label{omjer}
\end{equation}

\begin{figure}
\includegraphics{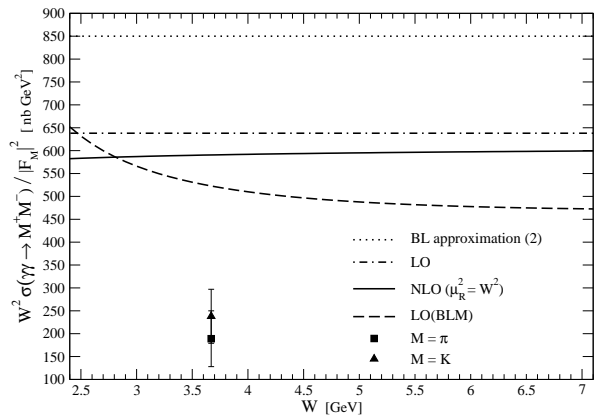}
\caption{\label{figomjer}  
Ratio $W^2\,
\sigma (\gamma\gamma \to M^+ M^-)/|F_M(W^2)|^2$.
The experimental points are obtained from the data of
\cite{Belle} and \cite{tlff}.}
\end{figure}

A comparison of this result with the only two reliable
experimental points, based on the data from \cite{Belle}
and \cite{tlff}, is given in Fig. \ref{figomjer}. The solid line 
corresponds
to Eq. (\ref{omjer}), with $\mu_R^2=W^2$, the dotted line
is the BL estimate following from Eq. (\ref{ffBL}), 
the dot-dashed line is the
complete LO prediction (equal to $638\,{\rm nb}\,{\rm GeV}^2$, 
according to Eq. (\ref{omjer})),
while the dashed
line is the ratio of the LO(BLM) results with the BLM
procedure performed separately for the cross section and
the form factor.
As it is seen, our results for the ratio are above the data by
approximately a factor of 2 to 3 which is in much better
agreement with the data than the
prediction for the cross sections based on Eq. (\ref{final}).
As Fig. \ref{figomjer} shows, the usage of the BLM scales
leads to much better agreement with the data. 
This, together with the fact that the LO(BLM) prediction lies 
inside the range of the NLO prediction 
(shaded area in Fig. \ref{figpion}), 
supports the idea behind the BLM procedure according to
which the BLM determined scales represent the relevant physical 
scales (putting aside understatements regarding the effective 
coupling).

In conclusion, the NLO results obtained for the process
$\gamma \gamma \to M^+ M^-$ ($M= \pi ,\,K$) are not conclusive enough
to allow a definite statement regarding the applicability
of the HSA at the currently accessible energies above 3 GeV.
However, improvements are possible in two directions:
the first, the more appropriate choice of the DA (possibly
in the direction suggested in Ref. \cite{Cher}), and the second,
a better understanding of the resummation for the
processes containng both spacelike and timelike scales
(possibly in the directions suggested in Refs. \cite{alphaV, 
Shirkov}).

\begin{acknowledgments}
The authors would like to thank S.~J.~Brodsky and N.~Kivel for
useful comments. 
\end{acknowledgments}


\end{document}